\documentclass[prl,twocolumn,floatfix,superscriptaddress]{revtex4-1}
\usepackage{dcolumn,amsmath}
\usepackage{graphicx}
\usepackage{bm}
\usepackage{hyperref}

\newcommand{\ecm}{\ensuremath{e {\cdotp} {\rm cm}}}
\newcommand{\eEDM}{{\em e}EDM}

\newcommand{\B}{\mathcal{B}} 
\newcommand{\E}{\mathcal{E}} 

\newcommand{\de}{d_\mathrm{e}}

\newcommand{\Dp}{D_{\|}}

\begin{document}
\title{Study of systematic effects in the HfF$^+$ ion experiment to search for the electron electric dipole moment}

\author{A.N.\ Petrov}\email{alexsandernp@gmail.com}
\homepage{http://www.qchem.pnpi.spb.ru}
\affiliation{National Research Centre ``Kurchatov Institute'' B.P. Konstantinov Petersburg Nuclear Physics Institute, Gatchina, Leningrad District 188300, Russia}
\affiliation{Saint Petersburg State University, 7/9 Universitetskaya nab., St. Petersburg, 199034 Russia}
\date{\today}

\begin{abstract}
The energy splittings for $J = 1$, $F=3/2$, $|m_F|=3/2$ hyperfine levels of the $^3\Delta_1$ electronic state of $^{180}$Hf$^{19}$F$^+$ ion are calculated as functions of the external variable electric and
magnetic fields within two approaches. In the first one transition to the rotating frame is performed, whereas in the second approach the quantization of rotating electromagnetic field is performed.
Calculations are required for understanding possible systematic errors in the experiment to search for electron electric dipole moment (eEDM) on $^{180}$Hf$^{19}$F$^+$ ion.
\end{abstract}

\maketitle

\begin{figure}
\includegraphics[width = 3.3 in]{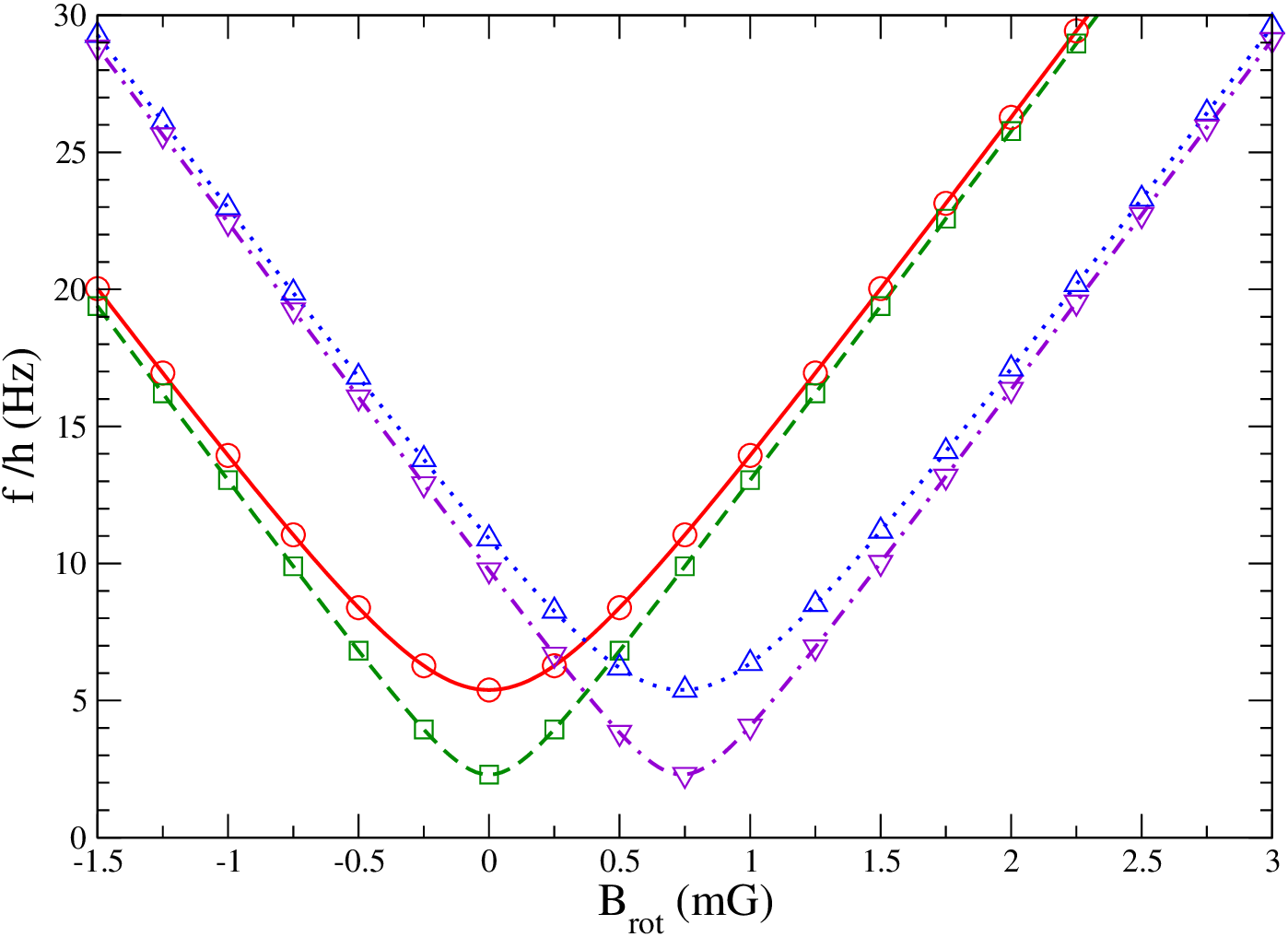}
\caption{(Color online) Calculated energy splittings for the $H^3\Delta_1$ ($J=1, F=3/2$, $|m_F|=3/2$) Stark pairs as functions of $\B_{\rm rot}$.
$\E_{\rm rot}=24$V/cm, $\omega_{\rm rot}/2\pi = 250$ kHz, $\tilde {\cal R}=+1$ in the calculations. Lines are calculated within approach (I). Solid (red) line corresponds to the lower ($\tilde {\cal D}=+1$) Stark pair, $\E_z=0$; dashed (green) line
corresponds to the upper ($\tilde {\cal D}=-1$) Stark pair, $\E_z=0$; Dotted (blue) line corresponds to the lower Stark pair, $\E_z=0.3$ mV/cm; Dotted-dashed (purple) line corresponds to the upper Stark pair, $\E_z=0.3$ mV/cm.
Figures are calculated within approach (II) with $N=3$. Circles (red) correspond to the lower Stark pair, $\E_z=0$; Squares (green)
correspond to the upper Stark pair, $\E_z=0$; up triangles (blue) correspond to the lower Stark pair, $\E_z=0.3$ mV/cm; down triangles (purple) correspond to the upper Stark pair, $\E_z=0.3$ mV/cm.}
 \label{avoid}
\end{figure}

Search for the electron electric dipole moment (\eEDM), $d_e$, is one of the most sensitive tests to-date for extensions of the standard model \cite{Commins:98,Chupp:15}. 
Very recently, Cornell/Ye group has obtained limit $|\de|<1.3\times 10^{-28}$ \ecm\ (90\% confidence)
using trapped $^{180}$Hf$^{19}$F$^+$ ions \cite{Cornell:17}. The result is in agreement with the best limit $|\de|<0.9\times 10^{-28}$ \ecm\ (90\% confidence) obtained in Ref. \cite{Baron2013}. 
The measurements were performed on the ground rotational, $J{=}1$, level in the metastable electronic $H^3\Delta_1$ state of $^{180}$Hf$^{19}$F$^+$ using the {\it rotating} electric and magnetic fields.
The \eEDM\ sensitive levels are described in details in Refs. \cite{Leanhardt:20111, Cornell:13, Cornell:17}. In brief, $^{180}$Hf isotope is spinless whereas $^{19}$F isotope has a non-zero nuclear spin $I{=}1/2$,
which gives rise to hyperfine energy splitting between levels with total, {\bf F=J+I}, momentum  $F=3/2$ and $F=1/2$. In the absence of external fields, each hyperfine level has two parity eigenstates known as the $\Omega$-doublet.
In the external rotating electric field the $F=3/2$ state splits to four Stark doublets levels. 
Two Stark doublets with projection of the total momentum
on the rotating field $m_F=\pm3/2$ are of interest for the \eEDM\ search experiment. The rotating magnetic field which is parallel or antiparallel to the rotating electric field further splits each Stark doublet to pair of Zeeman sublevels.
%
$m_F=\pm3/2$ sublevels are degenerate, in the absence of rotation, at zero magnetic field.
However, the rotation connects the sublevels and turns the degeneracy to a splitting at the avoided crossing between $m_F=+3/2$ and $m_F=-3/2$ sublevels (see Fig.(\ref{avoid})).

The energy splitting, $f$, between sublevels is measured in the experiment.
The measurement of $f$ is repeated under different conditions which can be characterized by three binary switch parameters $\tilde {\cal B}$, $\tilde {\cal D}$, $\tilde {\cal R}$ being switched from $+1$ to $-1$.
$\tilde {\cal B} =+1(-1)$ means that rotating magnetic field, ${\bf B}_{\rm rot}$, is parallel (antiparallel) to rotating electric field ${\bf E}_{\rm rot}$, $\tilde {\cal D}=+1(-1)$ means that the measurement was performed for lower (upper)
Stark level, $\tilde {\cal R}$ defines direction for the rotation of the fields. An \eEDM\ signal manifests as the main contribution to $f^{ {\cal B} {\cal D}}$ channel. Here notation $f^{S_1,S_2...}$ denotes a component which is odd under the switches $S_1,S_2,...$. The notations are close to those in Refs. \cite{ACME:17,Petrov:17c}. $f^{S_1,S_2...}$ can be calculated by formula
\begin{equation}
f^{S_1,S_2...} \left( \left| {\bf B}_{\rm rot} \right| \right) = \frac{1}{8}\sum_{\tilde {\cal B}, \tilde {\cal D}, \tilde {\cal R}}{S_1S_2...}f\left(   \B_{\rm rot}, \tilde {\cal D}, \tilde {\cal R} \right),
\label{components}
\end{equation}
where $\B_{\rm rot} = \tilde {\cal B}\left| {\bf B}_{\rm rot} \right| = \tilde {\cal B}\left| {\cal B}_{\rm rot} \right|$, ${S_1,S_2...}$ is a subset of the $\tilde {\cal B}$, $\tilde {\cal D}$, $\tilde {\cal R}$ parameters.
For simplicity, the only dependence of $f$ on 
$\B_{\rm rot}, \tilde {\cal D}, \tilde {\cal R}$ parameters is explicitly 
specified in eq. (\ref{components}).

The second-generation of \eEDM\ measurement experiment will provide an order of magnitude 
higher \eEDM\ sensitivity than the current limit \cite{Cornell:17}. It is rather clear though, that the 
increase in statistical sensitivity is only reasonable up to the level where 
systematic effects start prevailing. Thus accurate evaluation of systematic effects becomes
more important with the increase in statistical sensitivity. Such an analysis reduces 
to a theoretical study of different channels $f^{S_1,S_2...}$ as functions of electric and 
magnetic fields which is one of the goals of the present work.

One of the most important properties determining the prospects of molecules with 
regards to the search for \eEDM\ is the effective electric field, $E_{\rm eff}$,
which can be obtained only in the precise calculations of the electronic structure.
The \eEDM\ sensitive frequency,  $f^{ {\cal B} {\cal D}}$,  is proportional to both
$E_{\rm eff}$ and degree of polarization of molecule. Study of the latter as 
function of electric and  magnetic fields is the second goal of the paper.

 Following
Refs. \cite{Petrov:11,Petrov:14}, the energy levels and wave functions of the  $^{180}$Hf$^{19}$F$^+$ ion are obtained by numerical diagonalization of the molecular Hamiltonian (${\rm \bf \hat{H}}_{\rm mol}$) in external variable electric ${\bf E}(t)$ and magnetic ${\bf B}(t)$ fields 
over the basis set of the electronic-rotational wavefunctions
\begin{equation}
 \Psi_{\Omega}\theta^{J}_{M,\Omega}(\alpha,\beta)U^{\rm F}_{M_I}.
\label{basis}
\end{equation}
Here $\Psi_{\Omega}$ is the electronic wavefunction, $\theta^{J}_{M,\Omega}(\alpha,\beta)=\sqrt{(2J+1)/{4\pi}}D^{J}_{M,\Omega}(\alpha,\beta,\gamma=0)$ is the rotational wavefunction, $\alpha,\beta,\gamma$ are Euler angles, $U^{F}_{M_I}$ is the F nuclear spin wavefunctions and $M$ $(\Omega)$ is the projection of the molecule angular momentum, {\bf J}, on the lab $\hat{z}$ (internuclear $\hat{n}$) axis, $M_I=\pm1/2$ is the projection of the nuclear angular 
momentum on the same axis. Note that $M_F=M_I+M$ is not equal to $m_F$. The latter, as stated above, is the projection of the total momentum on the rotating electric field.

We write the molecular Hamiltonian for $^{180}$Hf$^{19}$F$^+$ in the form:
\begin{equation}
{\rm \bf\hat{H}}_{\rm mol} = {\rm \bf \hat{H}}_{\rm el} + {\rm \bf \hat{H}}_{\rm rot} + {\rm \bf\hat{H}}_{\rm hfs} + {\rm \bf\hat{H}}_{\rm ext} .
\end{equation} 
Here ${\rm \bf \hat{H}}_{\rm el}$ is the electronic Hamiltonian,
\begin{equation}
 {\rm \bf\hat{H}}_{\rm rot} = {\rm B}_0 { {\bf J}}^2 -2{\rm B}_0({ {\bf J}}\cdot{ {\bf J}}^e)
\end{equation}
is the Hamiltonian of the rotation of the molecule, ${\rm B}_0=0.2989$ cm$^{-1}$ \cite{Cossel:12} is the rotational constant. 
\begin{equation}
 {\rm \bf\hat{H}}_{\rm hfs} = {\rm g}_{\rm F}{\mu_{N}} {\bf \rm I} \cdot \sum_i\left(\frac{\bm{\alpha}_i\times \bm{r}_i}{r_i^3}\right)
\end{equation}
is the hyperfine interaction between electrons and flourine nuclei, ${\rm g}_F=5.25773$ is $^{19}$F nucleus g-factor,  $\mu_{N}$ is the nuclear magneton,

\begin{eqnarray}
\nonumber
 {\rm \bf\hat{H}}_{\rm ext}({\bf E}_{\rm static},{\bf B}_{\rm static},\E_{\rm rot},\B_{\rm rot}) = \\
\mu_{\rm B}({ {\bf L}}^e-{\rm g}_{S}{ {\bf S}}^e)\cdot{\bf B}(\rm t) -{\rm g}_{\rm F}\frac{\mu_{N}}{\mu_{B}}{\bf \rm I}\cdot{\bf B}(\rm t)   -{ {\bf D}} \cdot {\bf E}(\rm t)
\end{eqnarray}
describes the interaction of the molecule with external variable magnetic and electric fields. Here ${\rm g}_{S} = -2.0023$ is a free$-$electron $g$-factor, ${\bf J}^e = {\bf L}^e + {\bf S}^e$, ${\bf L}^e$ and ${\bf S}^e$ are the total electronic, electronic orbital and electronic spin momentum operators, respectively,
 {\bf D} is the dipole moment operator.
Variable fields are the sum of the static and rotating in the $xy$ plane components:
\begin{equation}
 {\bf E}(\rm t) =  {\bf E}_{\rm static} + {\bf E}_{\rm rot}(\rm t),
\label{Erot}
\end{equation}
\begin{equation}
 {\bf E}_{\rm rot}(\rm t) = \E_{\rm rot}(\hat{x}cos(\omega_{\rm rot}t) + \tilde {\cal R}\hat{y}sin(\omega_{\rm rot}t)),
\label{Erot2}
\end{equation}
\begin{equation}
 {\bf B}(\rm t) =  {\bf B}_{\rm static} + {\bf B}_{\rm rot}(\rm t) ,
\label{Brot}
\end{equation}
\begin{equation}
  {\bf B}_{\rm rot}(\rm t) = \B_{\rm rot}(\hat{x}cos(\omega_{\rm rot}t) + \tilde {\cal R}\hat{y}sin(\omega_{\rm rot}t)),
\label{Brot2}
\end{equation}
where $\tilde {\cal R} = \pm 1$ defines direction of rotation along the $\hat{z}$ axis: ${\vec{ \omega}}_{\rm rot} = \tilde {\cal R}\omega_{\rm rot}\hat{z}$.
$\tilde {\cal R} = + 1 (-1)$ if the fields rotate counter-clockwise (clockwise) around the $\hat{z}$ axis.
Below we put $\omega_{\rm rot}/2\pi = +250, +150$ kHz, $\E_{\rm rot}=+24, +20$ V/cm to values used in the experiment \cite{Cornell:17}.
Note, that $\omega_{\rm rot}$ and $\E_{\rm rot}$ are always positive.
In this paper the accounting for time dependence of external fields is performed by two approaches. In the first (or I below) approach transition to the rotating frame is performed:
\begin{align}
\nonumber
{\rm \bf\hat{H}}_{\rm mol}^{\rm I}={\rm \bf \hat{H}}_{\rm el} + {\rm \bf \hat{H}}_{\rm rot} + {\rm \bf\hat{H}}_{\rm hfs} + \\
\nonumber
{\rm \bf\hat{H}}_{\rm ext}({\bf E}_{\rm static}+\E_{\rm rot}\hat{x},{\bf B}_{\rm static}+\B_{\rm rot}\hat{x},\E_{\rm rot}=0,\B_{\rm rot}=0) \\
- \vec{\omega}_{\rm rot}\cdot { {\bf F}}.
\label{HamI}
\end{align}

In the second (or II below) approach the interaction with rotating fields
\begin{align}
\nonumber
\left( \mu_{\rm B}({ {\bf L}}^e-{\rm g}_{S}{ {\bf S}}^e) -{\rm g}_{\rm F}\frac{\mu_{N}}{\mu_{B}}{\bf \rm I} \right) \cdot{\bf B}_{\rm rot}(\rm t)   -{ {\bf D}} \cdot {\bf E}_{\rm rot}(\rm t) = \\
\nonumber
 (\B_{\rm rot}/2)\left( \mu_{\rm B}({ { L^e_{-\cal R}}}-{\rm g}_{S}{ { S^e_{-\cal R}}}) -{\rm g}_{\rm F}\frac{\mu_{N}}{\mu_{B}}{ \rm I_{-\cal R}} \right)e^{i\omega_{\rm rot}t} \\
\nonumber
- (\E_{\rm rot}/2){ D_{-\cal R}}e^{i\omega_{\rm rot}t} + \\
\nonumber
(\B_{\rm rot}/2)\left( \mu_{\rm B}({ { L^e_{+\cal R}}}-{\rm g}_{S}{ { S^e_{+\cal R}}}) -{\rm g}_{\rm F}\frac{\mu_{N}}{\mu_{B}}{ \rm I_{+\cal R}} \right)e^{-i\omega_{\rm rot}t}\\
 - (\E_{\rm rot}/2){ D_{+\cal R}}e^{-i\omega_{\rm rot}t}
\end{align}
 is replaced by the interaction with the corresponding quantized electromagnetic fields:
\begin{align}
\nonumber
{\rm {\bf\hat{H}}_{quant}} = \hbar\omega_{\rm rot}a^+a - \sqrt{\frac{2\pi\hbar\omega_{\rm rot}}{V}} \times \\
\nonumber
 \B_{\rm rot}\left( \mu_{\rm B}({ { L^e_{-\cal R}}}-{\rm g}_{S}{ { S^e_{-\cal R}}}) -{\rm g}_{\rm F}\frac{\mu_{N}}{\mu_{B}}{ \rm I_{-\cal R}} \right)a^+ \\
\nonumber
- \E_{\rm rot}{ D_{-\cal R}}a^+ +\\
\nonumber
\B_{\rm rot}\left( \mu_{\rm B}({ { L^e_{+\cal R}}}-{\rm g}_{S}{ { S^e_{+\cal R}}}) -{\rm g}_{\rm F}\frac{\mu_{N}}{\mu_{B}}{ \rm I_{+\cal R}} \right)a \\
 - \E_{\rm rot}{ D_{+\cal R}}a, 
\label{Hquant}
\end{align}
where $a^+$ and $a$ are photon creation and annihilation operators, $V$ is a volume of the system,
\begin{equation}
D_{\pm} = D_{x} \pm iD_{y}
\end{equation}
and the same is for other vectors.
To work with Hamiltonian (\ref{Hquant}) one need to add the quantum number $\left|{n} \right>$, where $n=\frac{V}{8\hbar \pi \omega_{rot}} \gg 1$ is number of photons.
The approach was developed in Ref. \cite{Petrov:15}. Then the total Hamiltonian in the approach (II) is
\begin{align}
\nonumber
{\rm \bf\hat{H}}_{\rm mol}^{\rm II}={\rm \bf \hat{H}}_{\rm el} + {\rm \bf \hat{H}}_{\rm rot} + {\rm \bf\hat{H}}_{\rm hfs} + \\
\nonumber
{\rm \bf\hat{H}}_{\rm ext}({\bf E}_{\rm static},{\bf B}_{\rm static},\E_{\rm rot}=0,\B_{\rm rot}=0) \\
+ \rm {\bf\hat{H}}_{quant}.
\label{HamII}
\end{align}
For the current study we have considered the following low-lying electronic basis states: $^3\Delta_1$,  $^3\Delta_2$,  $^3\Pi_{0^+}$ and $^3\Pi_{0^-}$.
Electronic matrix elements required to evaluate molecular Hamiltonian have been taken from Ref. \cite{Petrov:17b},
with exception for hyperfine structure constant $A_{\parallel}= -62.0~{\rm MHz}$ and dipole moment   $\Dp =-1.40~ {\rm a.u.}$ for $^3\Delta_1$
which have been taken from Ref. \cite{Cornell:17}.

Only the static fields parallel to ${\vec{ \omega}}_{\rm rot}$ ($\hat{z}$ axis)
are allowed in the first scheme, whereas the second approach is valid for arbitrary ${\bf E}_{\rm static},{\bf B}_{\rm static}$.  
 Including other rotating and oscillating fields with arbitrary directions and frequencies is also possible within approach (II).
However, working with the second approach one should ensure the convergence of the result with number, $N$, of photon states
$\left|{n_0 -N} \right>$, $\left|{n_0 -N + 1}\right>$, ...,  $\left|{n_0 - 1}\right>$,$\left|{n_0}\right>$,  $\left|{n_0+1}\right>$, ...,$\left|{n_0 +N - 1}\right>$, $\left|{n_0 +N}\right> $  included to the calculation.
In the absence of external fields and with the static fields $\E_z\hat{z}$, $\B_z\hat{z}$ (aligned along $\hat{z}$ axis) with sufficiently large number of photon states both approaches should give the same result.

In Fig. (\ref{avoid}) the calculated within two approaches $f\left(   \B_{\rm rot}, \tilde {\cal D}, \tilde {\cal R}=+1 \right)$ for $\tilde {\cal D}=+1$ and $\tilde {\cal D}=-1$ as functions of $\B_{\rm rot}$ are given.
Approach (II) with $N=3$ is in a complete agreement with approach (I). Adding $\E_z\hat{z}$ leads to the tilting of the rotating quantization axis away from the plane of rotation
by small angle $\E_z / \E_{\rm rot}$. This changes the accumulated Berry phase and shift the avoiding crossing from $\B_{\rm rot}=0$ point. The effect is described in details in Ref. \cite{Leanhardt:20111}.
Data from Fig.(\ref{avoid}) mean that for negative g-factor of $J=1, F=3/2$, positive $\B_{\rm rot}$, counter-clockwise rotation of ${\bf E}_{\rm rot}$ adding
of static electric field, ${\bf E}_{\rm static} = \E_z\hat{z}, \E_z > 0$, leads to decreasing of Zeeman energy splittings for the $H^3\Delta_1$ ($J=1, F=3/2$, $|m_F|=3/2$) Stark pairs.
This result confirms the theory of Ref. \cite{Cornell:13} used to determine
the sign for g-factor of $J=1, F=3/2$ from observed Zeeman energy splittings.

\begin{figure}
\includegraphics[width = 3.3 in]{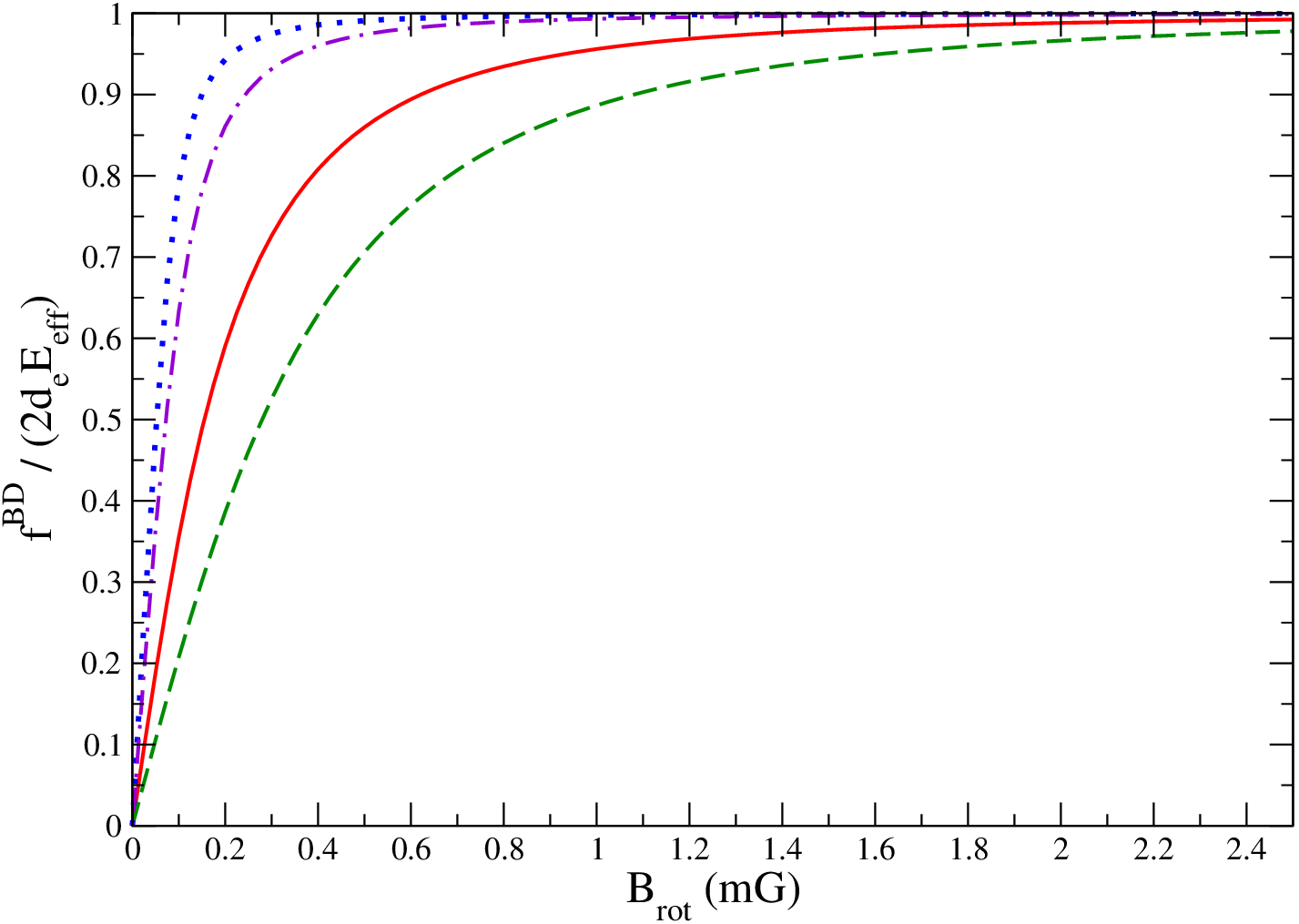}
\caption{(Color online) Calculated \eEDM\ induced $f^{ {\cal B} {\cal D}}$ splitting.
 Solid (red) line corresponds to $\omega_{\rm rot}/2\pi = 250$ kHz, ${\cal E}_{\rm rot} = 24$ V/cm;
dashed (green) line corresponds to $\omega_{\rm rot}/2\pi = 250$ kHz, ${\cal E}_{\rm rot} = 20$ V/cm;
dotted (blue) line corresponds to $\omega_{\rm rot}/2\pi = 150$ kHz, ${\cal E}_{\rm rot} = 24$ V/cm
dotted-dashed (purple) line corresponds to $\omega_{\rm rot}/2\pi = 150$ kHz, ${\cal E}_{\rm rot} = 20$ V/cm}
 \label{fBD}
\end{figure}

Interaction of \eEDM\ with the effective electric field $E_{\rm eff}=22.5~ {\rm GV/cm}$ \cite{Petrov:07a, Petrov:09b, Skripnikov:177} in the molecule
\begin{equation}
 {\rm \bf\hat{H}}_{\rm edm} = d_{e} E_{\rm eff} \left( \hat{n}\cdot {\bf J} \right)
\label{hedm}
\end{equation}
gives rise to $f^{ {\cal B}  {\cal D}}$ channel to be measured in the experiment.
To reach the maximum value $f^{ {\cal B}  {\cal D}} = 2d_{e} E_{\rm eff}$ laboratory electric field $\E_{\rm rot}$ must be large enough to
fully polarize molecule. $J=1$ HfF$^+$ becomes almost fully polarized  for  $\E_{\rm rot} > 5$ V/cm \cite{Cossel:12}.
%
However, the rotation causes the sublevels $m_F=+3/2$ and $m_F=-3/2$ to mix.
Therefore, at zero magnetic field, eigenstates are equal-mixed combinations of $m_F=\pm3/2$ sublevels which have different signs for \eEDM\ shift.
Thus value for magnetic 
field, $\B_{\rm rot}$, has also to be large enough to saturate $f^{ {\cal B}  {\cal D}}$ at $2d_{e} E_{\rm eff}$.
 In Fig. (\ref{fBD}) the calculated $f^{ {\cal B} {\cal D}}$ as a function of $\left| {\cal B}_{\rm rot} \right|$
are given. Both methods are in agreement.
Value for rotating magnetic field is given by \cite{Cornell:17} 

\begin{equation}
{\cal B}_{\rm rot} = {\cal B}^{\prime}_{\rm axgrad}r_{\rm rot},
\label{Brott}
\end{equation}
where ${\cal B}^{\prime}_{\rm axgrad} = 40$ mG/cm,  
\begin{equation}
r_{\rm rot} = \frac{e\E_{\rm rot}}{M\omega^2_{\rm rot}}
\label{rrot}
\end{equation}
 is the ion's radius of circular motion, $M = 199$ amu is mass of HfF$^+$.
For $\omega_{\rm rot}/2\pi = 250$ kHz eqs. (\ref{Brott},\ref{rrot}) give ${\cal B}_{\rm rot} = 1.87$ and ${\cal B}_{\rm rot} = 1.56$ G
 for ${\cal E}_{\rm rot} = 24$ and ${\cal E}_{\rm rot} = 20$ V/cm respectively.
Then, according to Fig. (\ref{fBD}), 98.5\% and 95\% efficiency is reached for ${\cal E}_{\rm rot} = 24$ V/cm and ${\cal E}_{\rm rot} = 20$ V/cm
which corresponds to effective electric field $E_{\rm eff}=22.2$ and $E_{\rm eff}=21.3~ {\rm GV/cm}$ respectively.
Note, that $E_{\rm eff}$ can not be measured but it is required for extracting EDM value from measured $f^{BD}$. 
See eqs. (3,4) in \cite{Cornell:17}.

\begin{figure}
\includegraphics[width = 3.3 in]{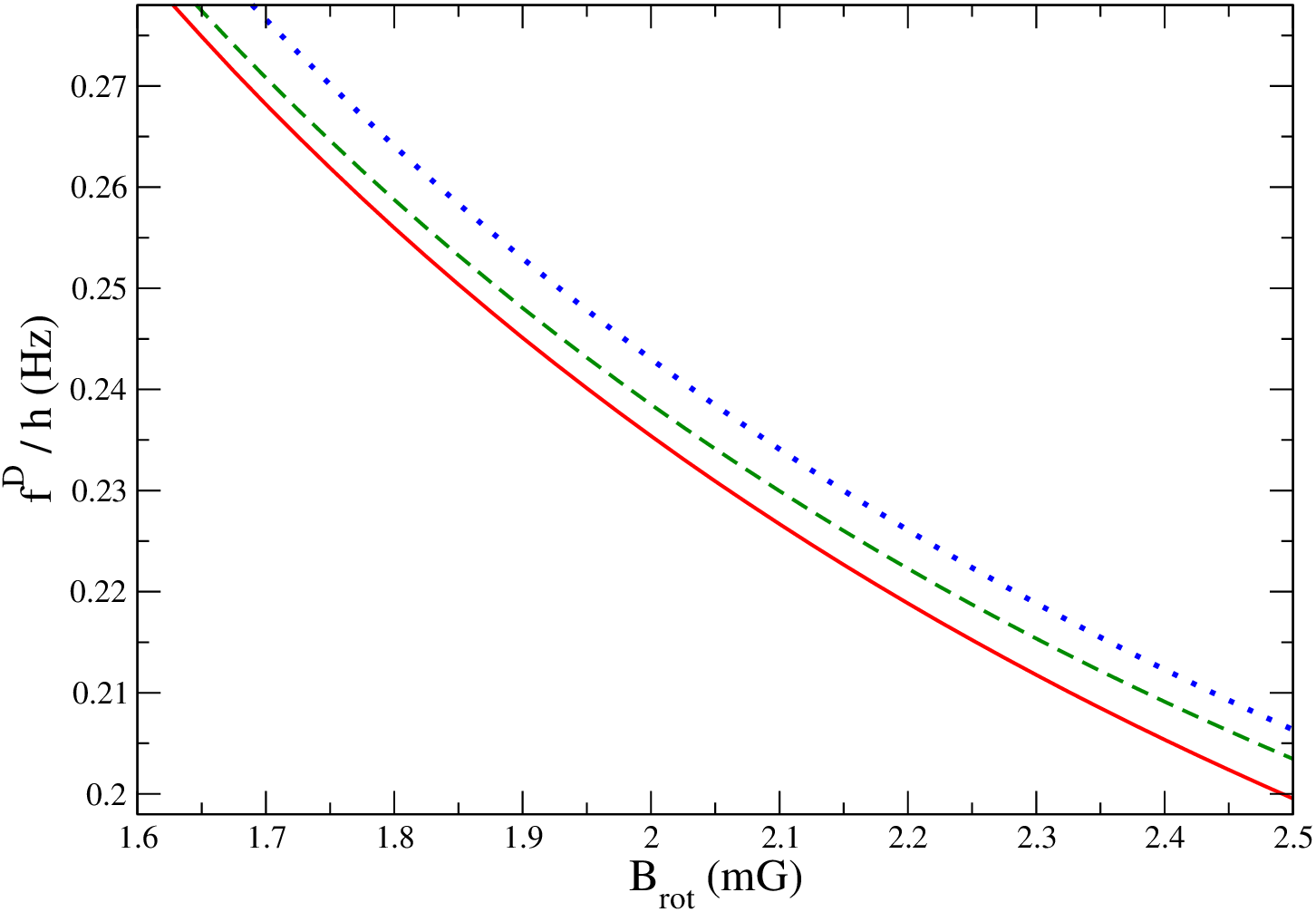}
\caption{(Color online) Calculated $f^{ {\cal D}}$ as function of ${\cal B}_{\rm rot}$.
Solid (red) curve: Interactions with both $^3\Delta_2$ and  $^3\Pi_{0^\pm}$ 
states are taken into account. Dashed (green) curve: Only interactions with the $^3\Pi_{0^\pm}$ states
are taken into account. Dotted (blue) curves: Interactions with both $^3\Delta_2$ and  $^3\Pi_{0^\pm}$ 
states are omitted. $\E_{\rm rot}=24$V/cm, $\omega_{\rm rot}/2\pi = 250$ kHz in the calculations.}
 \label{FD}
\end{figure}

\begin{figure}
\includegraphics[width = 3.3 in]{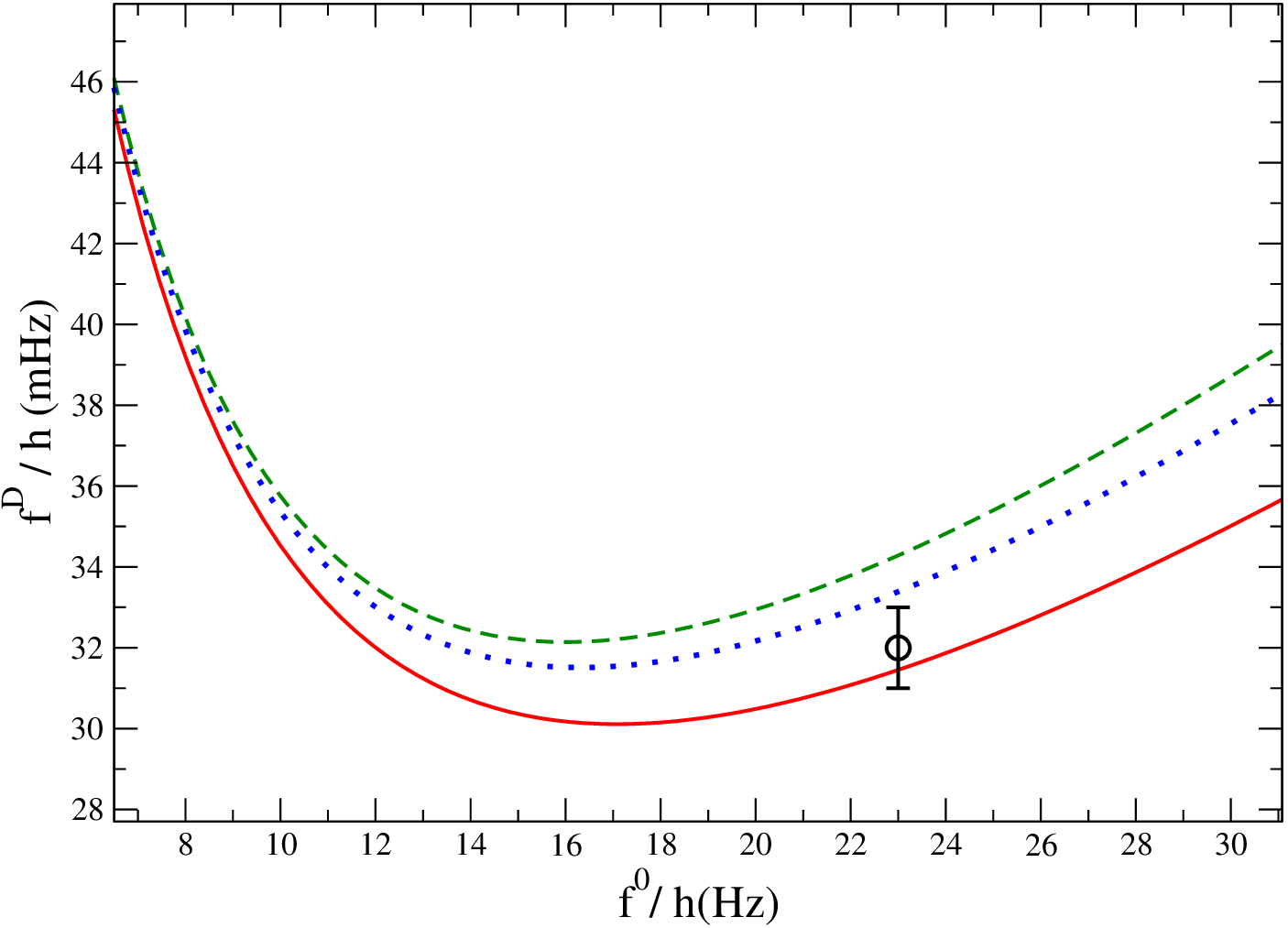}
\caption{(Color online) Calculated $f^{ {\cal D}}$ as function of $f^{0}$.
Solid (red) curve: Interactions with both $^3\Delta_2$ and  $^3\Pi_{0^\pm}$ 
states are taken into account. Dashed (green) curve: Only interactions with the $^3\Pi_{0^\pm}$ states
are taken into account. Dotted (blue) curves: Interactions with both $^3\Delta_2$ and  $^3\Pi_{0^\pm}$ 
states are omitted. $\E_{\rm rot}=24$V/cm, $\omega_{\rm rot}/2\pi = 150$ kHz in the calculations. Circle is the experimental datum.}
\label{FD2}
\end{figure}

One of the main systematic effect in the experiment for \eEDM\  search on $^{180}$Hf$^{19}$F$^+$ ions comes from doublet population contamination 
(population of lower (upper) Stark doublet when only upper (lower) one should be populated) \cite{Cornell:17}. The extent of the contamination is
estimated from difference between measured and predicted (calculated) values of $f^{ {\cal D}}$.
In Fig. (\ref{FD}) the calculated $f^{ {\cal D}}$ as function of ${\cal B}_{\rm rot}$ for $\omega_{\rm rot}/2\pi = 250$ kHz, ${\cal E}_{\rm rot} = 24$ V/cm
is given. One sees that accounting for interaction with $^3\Delta_2$,  $^3\Pi_{0^+}$ and $^3\Pi_{0^-}$ electronic states is important
for accurate calculation of $f^{ {\cal D}}$ and change result on about 4 \%.
In Fig. (\ref{FD2}) the calculated $f^{ {\cal D}}$ as function of $f^{ 0}$ and experimental
value \cite{Cornell:17} $f^{0}/{ h} =$ 22.9985(13) Hz, $f^{ {\cal D}}/{ h} =$ 32.0(1.0) mHz for $\omega_{\rm rot}/2\pi = 150$ kHz, ${\cal E}_{\rm rot} = 24$ V/cm are given. To plot Fig. (\ref{FD2}) both
$f^ {\cal D}$ and $f^{0}$ are assumed to be functions of  ${\cal B}_{\rm rot}$. One sees that accounting
for the contribution of interaction with $^3\Delta_2$,  $^3\Pi_{0^+}$ and $^3\Pi_{0^-}$ electronic states leads 
to agreement between the measured and calculated values.

Finally, we have calculated the effective electric field  $E_{\rm eff}$ and energy splittings for $J = 1$, $F=3/2$, $|m_F|=3/2$ hyperfine levels of the $^3\Delta_1$ electronic state 
as functions of the external electric and magnetic fields. It is shown that for accurate evaluation of $f^{ {\cal D}}$ frecuency the interaction with 
$^3\Delta_2$,  $^3\Pi_{0^+}$ and $^3\Pi_{0^-}$ electronic states has to be taken into account. Calculation of $f^{ {\cal D}}$ is required
for estimation of systematic effect related with doublet population contamination.

The work is supported by the Russian Science Foundation grant No. 14-31-00022.


\end{document}